%% file: main.tex
\documentclass[12pt,a4paper]{article}

\input{preamble.tex}
\usepackage{fullpage}
\usepackage[ruled,vlined]{algorithm2e}

\usepackage{authblk}

\author[1, 2]{Ittai Rubinstein}

\affil[1]{Blavatnik School of Computer Science, Tel-Aviv University, Tel-Aviv 69978, Israel}
\affil[2]{QEDMA Quantum Computing, Tel-Aviv, Israel}

\title{Explicit and Efficient Construction of (nearly) Optimal Rate Codes for the BDC and PRC Channels}

\date{\today}

\begin{document}
\maketitle

\begin{abstract}
Two of the most common models for channels with synchronisation errors are the Binary Deletion Channel with parameter $p$ ($\text{BDC}_p$) -- a channel where every bit of the codeword is deleted i.i.d with probability $p$, and the Poisson Repeat Channel with parameter $\lambda$ ($\text{PRC}_\lambda$) -- a channel where every bit of the codeword is repeated $\text{Poisson}(\lambda)$ times.

Most previous constructions based on synchronisation strings yielded codes with rates far lower than the capacities of these channels \cite{CS18,guruswami2018polynomial}, and the only efficient construction to achieve capacity on the BDC at the time of writing this paper is based on the far more advanced methods of polar codes \cite{tal2021polar}.

In this work, we present a new method for concatenating synchronisation codes and use it to construct simple and efficient encoding and decoding algorithms for both channels with nearly optimal rates.
\end{abstract}

\section{Introduction}
The theory of error-correcting--codes deals with methods for encoding messages to be sent over noisy media in such a manner that they can be correctly decoded afterwards. 
Initially introduced by Shannon \cite{shannon1948mathematical}, this field has proven to be instrumental in understanding the theory of computation, and has had a  wide variety of applications in other fields, such as communications, and computational biology \cite{vanstone2013introduction}.

The most commonly considered models are “Synchronous Models” - models where the message may be altered or erased, but every letter that was received can be traced back to its original position in the transmitted message. 
This category includes models such as the Binary Symmetry Channel (BSC) where some of the bits in the transmitted message are flipped (i.e. changed from 1 to 0 or vice versa), and the Binary Erasure Channels (BEC) where some of the bits of the transmitted message are replaced with a question mark (but are not removed, thus preserving the alignment between the transmitted message and the received message). 
These models can be adversarial (such as \cite{hamming1950error}), where the code must correct any error the channel may produce, or average-case (such as \cite{shannon1948mathematical}), where the effect of the channel is random and decoding only needs to succeed w.h.p. 

Synchronous models are very well researched, and have a variety of efficient encoding and decoding algorithms \cite{vanstone2013introduction}. 
The main method used for such channels are linear error-correcting--codes and they rely heavily on the fact that the letters of the received messages can be mapped back into letters of the transmitted messages. 

However, in many real world applications, deletions and insertions can cause the received codewords to be misaligned and the decoder must also deal with synchronisation errors \cite{mitzenmacher2009survey, mercier2010survey}. 
Perhaps the most intuitive asynchronous channel is the Binary Deletion Channel. 
This is a channel where every transmitted bit is deleted i.i.d with probability $p$ (where $1 > p > 0$ is a parameter of the channel). 

Here, unlike with the binary erasure channel, deleted bits are not replaced with a question mark, but are completely removed from the sequence, shifting the rest of the received codeword.
For instance, when transmitting the codeword $w=(1, 0, 0, 0, 1, 1)$, the BEC may result in the received codeword $r_\text{BEC}= (1, 0, ?, ?, 1, ?)$ while a similar pattern of deletions would result in the received codeword $r_\text{BDC} = (1, 0, 1)$.

This channel represents a simple model for many real-life systems in which there is a loss of information due to synchronisation errors. 
Moreover, the tools developed for this channel have been instrumental in a variety of other fields \cite{mitzenmacher2009survey, mercier2010survey}.

A slightly more complex model which we will also consider, is the Poisson Repeat Channel with parameter $\lambda > 0$ ($\text{PRC}_\lambda$).
This is a channel where every transmitted bit is received $\text{Poisson}(\lambda)$ times. 
While originally used to help prove lower bounds on the capacity of binary deletion channels, Poisson repeat channels are interesting in their own right.
Indeed, Poisson repeats can model every-day examples like a sticky key in a keyboard, as well as deeper technological issues such as errors in single photon generation (a crucial step in light-based quantum computing) \cite{buller2009single}.

However, the PRC also presents a slightly greater challenge.
This is because bits can now be received more than once, resulting in several new types of synchronisation errors.
Continuing with our previous example where the codeword $w=(1, 0, 0, 0, 1, 1)$ was transmitted, the received codeword in the BDC model will always start with at most a single $1$ bit, unless all three $0$s were deleted.
In the Poisson repeat channel, this is not the case and the received codeword $r_\text{PRC} = (1, 1, 0, 1)$ is a possible outcome.

Several previous results have shown an interesting connection between these two noise models.
For instance, Mitzenmacher and Drinea's lower bound for the capacity of the BDC channel \cite{MD06} is based on their previous lower bound for the capacity of the PRC channel, and Con and Shpilka's constructive codes for the BDC channel \cite{CS18} are also applicable to the PRC channel.

In this paper we will focus our attention on these two channels, but we believe our tools and approaches can be applicable to other asynchronous channels as well.

\subsection{Previous Work}

Asynchronous channels present us with a varied field of research, and we will not be able to cover all of its results here.
The excellent surveys by Mitzenmacher, Cheraghchi et al and Mercier et al \cite{mitzenmacher2009survey, mercier2010survey, cheraghchi2020overview} give a more detailed background.

Determining the capacity of the $\text{BDC}_p$ channel, remains an open problem, and so far it has been answered only for some extremal cases.
When $p\rightarrow 0$, the capacity of this channel is $1-h(p)$ \cite{kalai2010tight} (where $h(\cdot)$ is the binary entropy function), when $p \rightarrow 1$, the capacity is $\mu (1 - p)$ where $\frac{1}{9} < \mu \leq 0.4143 - o(1)$ \cite{MD06, dalai2011new} (where $o(1)$ is w.r.t the block size $n$), and \cite{venkataramanan2013achievable} give lower bounds for some of the intermediate values of $p$.

However, the lower bound on the capacity by Mitzenmacher and Drinea \cite{MD06} is not based on an efficient construction.
Recently, Con and Shpilka constructed a family of error-correcting-codes with rate $r \geq \frac{1-p}{16}$ for the $\text{BDC}_p$ channel and $r \geq \frac{\lambda}{17}$ for the $\text{PRC}_\lambda$ (when $\lambda < \frac{1}{2}$) \cite{CS18}, improving upon the results of Guruswami and Li who presented the first explicit codes with rate $\Theta(1-p)$ in \cite{guruswami2018polynomial}.
This was improved upon by, Tal et al and by Pfister and Tal who showed that polar codes can be used to construct efficiently decodable codes with optimal rates for the BDC channel \cite{tal2021polar, pfister2021polar}.

Haeupler and Shahrasbi \cite{haeupler2017synchronization} construct a family of efficient codes for InsDel channels with a sufficiently large alphabet, and Haeupler, Rubinstein and Shahrasbi improve the decoder in \cite{haeupler2019near}, reaching quasi-linear complexity.

In the adversarial model, Guruswami and Wang \cite{guruswami2017deletion} showed that there are codes with rate $1 - \widetilde{O} \left(\sqrt{\delta}\right)$ that can correct up to $\delta n$ errors.
This rate was improved by Cheng et al. \cite{cheng2018deterministic} and further by Haeupler \cite{haeupler2019optimal}.
More recently it was shown by Con, Shpilka and Tamo \cite{con2021linear} that, surprisingly, linear error correcting codes are also effective for adversarial InsDel channels.

\subsection{Main Contribution}

In this work we will construct a family of codes, with efficient encoding and decoding algorithms, whose rates are arbitrarily close to the capacities of the BDC and PRC channels.
Unlike the previous results of Tal et al. and of Pfister and Tal \cite{tal2021polar, pfister2021polar}, this construction does not require the more advanced machinery of polar codes.
Furthermore, the construction presented has a decoding error probability of $\exp{\left(-\Theta\left(n^\frac{1}{6}\right)\right)}$ (where $n$ is the block length) with a quasi-linear complexity decoder, while Pfister and Tal's code requires $n^{\frac{3}{2} + \varepsilon}$ time for the same error probability.

Both our code and Pfister and Tal's construction assume that we are given some inner code which achieves a high rate on the channel but which does not necessarily have efficient encoding and decoding algorithms, and both methods produce a new version of this code with efficient encoding and decoding.
However, Pfister and Tal's construction requires this code to be generated by a hidden-Markov distribution, while the construction presented here can be used with any inner code.
Li et al. proved that there exists a finite hidden-Markov distribution code that achieves capacity for this channel \cite{li2020capacity} and this model can clearly be found in $O(1)$ time using an enumeration technique similar to the one described in Section \ref{subsec:base}, but such a code has not been found yet.

\begin{theorem} [Main Result (informal)]
Any (possibly inefficient) family of codes for either the $BDC$ or the $PRC$ channel can be converted into a family of codes for the same channel with an arbitrarily close rate, that has encoding and decoding algorithms with a quasi-linear complexity.
\end{theorem}

It should be noted that while this complexity is asymptotically very good, it hides within it a very large constant factor, and while we do not have an exact bound on it, we expect it to grow as the rate of the code approaches the capacity of the channel.

Nonetheless, this allows us to construct to construct a family of efficient codes which achieve rates of $\frac{1-p}{9}$ for the $\text{BDC}_p$ channel, thus completing the line of works started by Guruswami and Li \cite{guruswami2018polynomial} and continued by Con and Shpilka \cite{CS18}.

Our construction is based on a new technique for tracking inner codewords in a concatenation of synchronisation codes, which we hope may be useful in other cases as well:

Most previous constructions use buffers of $0$s as the delimiters between inner codewords.
Then, by bounding the probability that the channel would transform any substring of the inner codeword into a sufficiently long sequence of $0$s, they can ensure that long sequences of $0$s in the received codeword mostly correspond to delimiters. 
In other words, when separating the received codeword into inner codewords, one searches the entire string for patterns that may have come from a delimiter.
    
In our construction, we will use delimiters in a very different manner.
Instead of searching the entire codeword for the delimiters, we will use our knowledge of the length of the inner codeword and the average expansion of the channel to produce a prior estimate for the distance between consecutive delimiters.
Using this prior estimate, we are able to find the delimiters one after another, without looking at the entire codeword.

This new method allows us to drastically reduce the probability that even a single inner delimiter will be missed, while reducing the overhead of the delimiters to a negligible fraction of the codeword.
The ability to decode under the assumption that all delimiters will be found simplifies the outer code in our construction, and the fact that we will not search for a delimiter within an inner codeword allows us to use a general inner code, resulting in a nearly optimal rate.

\subsection{An Overview of Con and Shpilka's Construction}
Since our approach will be similar to that of \cite{CS18}, we will begin with a short overview of their construction, which is based on a concatenation of codes. 

Initially, the message is divided into segments of length $\sigma = O(1)$. 
These are thought of as members of an alphabet $\Sigma$ of size $\vert \Sigma \vert = 2^\sigma$ and can be encoded using \cite{haeupler2017synchronization}. 
Each letter in the encoded message is then converted back to a string of $\sigma$ binary symbols and is encoded using an inner code. 
Since the inner code is only applied to strings of length $\sigma=O(1)$, it can be inefficient without affecting the asymptotic complexity of the encoding / decoding algorithms. 
The encoded strings are appended and separated by delimiters - in this case buffers of 0s.

Con and Shpilka's construction has a quasi-linear encoding algorithm and a quadratic decoding algorithm, where the computational bottleneck of the decoding algorithm comes from decoding the outer code.
The improved decoding algorithm for Haupler and Shahrasbi's code \cite{haeupler2019near} can be used with Con and Shpilka's code to reduce the complexity of their decoding algorithm to a quasi-linear time as well.

\subsection{Sketch of the Proof}
Our construction will be based on a similar strategy, but with a few key differences.
Firstly, we note that most of the overhead of this code is caused by the fact that the inner code is designed to preserve a certain structure. 

By removing this structure we are able to significantly increase the rate of our code.
However, this comes at a cost - separating unstructured codewords from the delimiters is made far more difficult. 
We overcome this using a slightly more complex construction of delimiters and a careful analysis.

In addition, the delimiters themselves account for another constant fraction of the overhead of Con and Shpilka's code.
By using a recursive concatenation, we are able to reduce the cost of these delimiters to a negligible fraction of the overhead.

At each step of this recursion, we will assume that there exists a BDC/PRC code with message length $k$ and block length $n$, and we will construct a code with message length $k ^ 2$ and block length $n ^ \prime = \left(1 + o(1) \right)nk$.

We will do this by separating the $k^2$ bit message into $k$ strings of $k$ bits each.
We will think of each of these $k$-bit strings as a member of an alphabet of size $\vert \Sigma \vert = 2 ^ k$ and use a $\text{ReedSolomon} \left[k+2t, k, t\right]_{2^k}$ code to give it some redundancy (i.e. a Reed-Solomon code over a field of size $2^k$, message length $k$ and block length $k+2t$ with distance $t = o(k)$).

If we were constructing a code for a discrete memoryless channel (DMC) such as the BEC, this step might not be very surprising, because without synchronisation errors we could map the received codeword back into the letters of the Reed-Solomon codeword.
However, in our case this might seem somewhat counter-intuitive, since Reed Solomon codes offer no protection against the synchronisation errors we are trying to correct. 
This step works for asynchronous channels as well because our delimiters are designed to fully preserve the synchronisation {\em between} inner codewords and the Reed Solomon code will only need to compensate for a small number of local decoding failures of inner codewords.

We will encode each of the $k+2t$ letters of the Reed Solomon codeword using our inner code. 
This will output a list of $k+2t$ strings of length $n$ bits each. 
Finally, we will append these strings after inserting a delimiter between each two.

The delimiters will be made up of two parts: a positioning string which will help us find the delimiter and two partitioning strings which will help us separate between the delimiters and the inner codewords themselves. 
The reason that we need partitioning strings, is that we make no assumptions about the structure of the inner code.
Therefore, any sequence of bits that originated from the delimiter could have originated from the inner codeword.

For instance, suppose we had used buffers of 0s as our delimiters.
Since we make no assumptions about the structure of the inner code, we have no way of knowing whether or not the inner codeword begins with a sequence of 0s, so we can't tell where the delimiter ends and the inner codeword begins.

This makes separating the two a very difficult task and will be at the heart of our construction.
Our separation between inner codewords and delimiters will not be completely accurate, but we will be able to bound the effect this has on the decoding failure probability by using the fact that the inner code is designed to deal with (some) deletions.

Both parts of the delimiter will be based on an idea we call “valleys”. 
Similar to the markers defined by Cheraghchi et al \cite{cheraghchi2020coded}, we will define valleys to be a long sequence of 0s followed by a long sequence of 1s (when looking at the cumulative sum of the string minus $\frac{1}{2}$, these translate to local minima - see Figure \ref{fig:align_alg}). 
Unless one of these two sequences is completely deleted by the channel, a valley in the transmitted message will result in a valley in the received message.

We will use this observation to align indices within the received message to their source in the transmitted message.
We will start with an estimate of where the center of some valley from the transmitted message should be in the received message, and then we will go downhill to the nearest local minima (see Algorithm \ref{alg:align_valley}).
By bounding both the probability that one of these sequences was removed and the probability that our initial estimate was outside the bounds of the received valley, we can correlate the center of the received valley to the center of the transmitted valley with high probability.

Each positioning string will be a long valley, and each partitioning string will be a short valley. 
We set the positioning string to be long, because we need to be able to find its center, given only a very rough estimate.
On the other hand, setting the partitioning strings this long would reduce the accuracy of its separation from the inner codewords.

The decoding algorithm will be similar to the encoding algorithm, but in a reversed order. 
First, we will align the received message by locating the centers of the positioning strings. 
Then we will use the partitioning strings to separate the delimiters from the inner codewords, and apply the inner code decoding to obtain the inner codewords. 
Finally, we use the Reed Solomon decoding to correct up to $t$ errors that may have occurred.

\subsection{Organization}
In Section \ref{sec:prelim} we will define basic notations, show some well-known inequalities that we will use in our analysis and present the basic building block of our construction. 
Section \ref{sec:recursion} contains the construction of our recursive step and in Section \ref{sec:base} we will prove the basis of the recursion and show how we can connect it to the recursive step. 
In Section \ref{sec:PRC} we will extend our results to the Poisson repeat channel, and in Section \ref{sec:analysis} we will bound the decoding failure rates for both channels. 
Finally, in Section \ref{sec:discussion} we will discuss the implications and limitations of these results, as well as potential avenues for future research.

\section{Preliminaries}
\label{sec:prelim}

\subsection{Average-case Codes}

\begin{definition}
    Let $\Sigma$ be a finite set and let $k, n\in\mathbb{N}$ be positive integers.
    
    We will say that $C$ is a {\em random channel} acting on the {\em alphabet} $\Sigma$ and {\em block-length} $n$ if it maps any member of $\Sigma^n$ to a distribution on some set $\mathcal{Y}$.
    
    Furthermore, we will say that {\em encoding and decoding algorithms} $E: \Sigma^k \rightarrow \Sigma^n, D:\mathcal{Y} \rightarrow \Sigma^k$ for this channel with {\em message length} $k$ have {\em rate} $\rho = \frac{k}{n}$ and a {\em decoding falure probability (DFP)} of
    \[
        \delta = \max_{m\in \{0,1\}^k} \left\{ \Pr\left[ D(C(E(m))) \neq m\right] \right\}
    \]
\end{definition}

In other words, the DFP of a code is the probability that a message will be decoded incorrectly if the message was chosen adversarially, but the effects of the channel were random.

\begin{definition}
    Let $C$ be a random channel. 
    We will say that $\mathcal{F} = \left\{\left(E_i, D_i\right)\right\}_{i\in\mathbb{N}}$ is a family of codes for $C$ if:
    \begin{itemize}
        \item The message lengths $k_i$ of $E_i, D_i$ are unbounded ($k_i \xrightarrow[i\rightarrow \infty]{} \infty$)
        \item The DFPs $\delta_i$ of $E_i, D_i$ in $C$ are vanishing ($\delta_i \xrightarrow[i\rightarrow \infty]{} 0$)
    \end{itemize}
\end{definition}

Throughout the decoding process we will often attempt to align the received message with the transmitted codeword.

\begin{definition}
When the channel acts independently on each letter of the input (i.e. when $C\left(b_1 , \ldots, b_n\right) = C\left(b_1\right)\ldots C\left( b_n\right)$), we will say that the $i$th coordinate of a message transmitted over some asynchronous channel and the $j$th coordinate of the received message are {\em aligned}, if the first $i-1$ letters of the transmitted message were mapped to a string of length at most $j$ by the channel and the first $i$ letters of message were mapped to at least $j$ letters by the channel.
\end{definition}

\subsection{Probability Inequalities}

Throughout this paper we will bound the probability that several parts of our construction will fail.
This will require several bounds on the tails of Poisson and binomial distributions, which we will present in this section.
Perhaps the most important tool at our disposal will be the Chernoff bound.

\begin{theorem} [Chernoff Bound for Binomial and Poisson Distributions]
\label{thm:chernoff}
    For all $\frac{1}{2} > \varepsilon, p > 0$, $n\in\mathbb{N}$ and $\lambda > 0$:
    \[
    \Pr \left[ \left\vert\text{Bin}\left(p, n\right) - p n \right\vert > \varepsilon p n \right] \leq 2 \exp \left( - \frac{1}{4} \varepsilon ^ 2 p n \right)
    \]
    \[
    \Pr \left[ \left\vert\text{Poisson}\left(\lambda\right) - \lambda \right\vert > \varepsilon \lambda \right] \leq 2 \exp \left( - \frac{1}{4} \varepsilon ^ 2 \lambda \right)
    \]
\end{theorem}

This theorem is based on the Chernoff bound (see Section 2.2 of \cite{boucheron2013concentration}) and an analysis of entropy functions.
Its proof will be shown in Appendix \ref{app:chernoff}.

Another special case of the Chernoff bound which we will prove in Appendix \ref{app:chernoff2} is

\begin{theorem} [Binomial Tail]
\label{thm:chernoff2}
    For all $\alpha > e^2$, $\frac{1}{\alpha+1} > p > 0$ and $n\in\mathbb{N}$:
    \[
    \Pr \left[ \text{Bin}\left(p, n\right) > (\alpha + 1) p n \right] \leq \exp \left( - \frac{1}{2} \log\left(\alpha\right) \alpha p n \right)
    \]
\end{theorem}

\subsection{Aligning Valleys}
Our main tool in aligning the messages will be an idea we call valleys.

\begin{definition}
    We will define a {\em valley} with {\em faces} of length $x$ and $y$ to be a string of $x$ 0s followed by $y$ 1s.
    
    In other words $\text{Valley}(x,y) = 0^x 1^y$.
    
    We will define the {\em center} of a valley to be the index where it transitions from $0$s to $1$s (i.e. the index of the last $0$ bit).
\end{definition}

Suppose the codeword contains some valley $V = \text{Valley}(x,y)$.
This is encoded by the channel into a valley whose face lengths are i.i.d $\text{Bin}(1-p, x)$ and $\text{Bin}(1-p, y)$ (or $\text{Pois}(\lambda, x)$ and $\text{Pois}(\lambda, y)$ for the PRC channel).
Assuming neither side of the valley was completely deleted, the center of the received valley and the center of the transmitted valley should be aligned.

Our definition of valleys is similar to the markers considered in \cite{cheraghchi2020coded}, but we will view our valleys very differently.
Cheraghchi et al construct their inner code so that is is unlikely to contain a long sequence of $0$s.
This allows them to find the delimiters by searching for the next sequence of $0$s and then use the marker to zero in on the center of the delimiter.

In our decoding algorithm, we will produce an approximate estimate for where the center of a certain valley should be, and by finding its actual center we will be able to align that index to a position within the transmitted message.
We will do so using Algorithm \ref{alg:align_valley}.

Intuitively, this algorithm works by using the observation that if the valley was not distorted by the channel too much and our initial estimate for its center was not off by too much, then we known whether or not our guess was to the right or to the left (since the left face of the valley has $0$s and the right face has $1$s).
We can then correct our position. 
If we are on the left face, go right until we reach the center and vice versa.

\begin{algorithm}[H]
\SetAlgoLined
\label{alg:align_valley}
\SetKwInOut{KwIn}{Input}
\SetKwInOut{KwOut}{Output}
    \KwIn{a received codeword $w\in{\{0,1\}}^*$, estimated center of valley $j$}
    \KwOut{the center of the valley $j^\prime$}
    $j ^ \prime \leftarrow j$\;
    \eIf{$w[j] = 0$}{
        \While{$w[j^\prime] \neq 1 $}{
            $ i \leftarrow i + 1 $ \;
        }
        \KwRet $j^\prime - 1$ \;
    }{
        \While{$w[j^\prime] \neq 0 $}{
            $ i \leftarrow i - 1 $ \;
        }
        \KwRet $j^\prime$ \;
    }
    \caption{Align Valley}
\end{algorithm}

In order to clarify this approach we will consider a simple example.
Suppose we are building a code for the $\text{BDC}_p$ channel with parameter $p = \frac{1}{2}$ and we decide to use a valley of length $32$ on either side to align some index.

That is, our transmitted message would be as follows (where the bold digit signifies the center of the valley).
\[
m = 
[\ldots] 0000000000000000000000000000000{\mathbf 0}11111111111111111111111111111111 [\ldots]
\]

Suppose the channel made the following deletions:

\[
d = [\ldots] \cdot\cdot\cdot0\cdot\cdot\cdot\cdot\cdot\cdot000\cdot0\cdot\cdot00\cdot\cdot\cdot\cdot0\cdot0\cdot0\cdot\cdot0\cdot\cdot1\cdot1\cdot11\cdot11\cdot1\cdot\cdot\cdot111\cdot1111111\cdot\cdot\cdot\cdot\cdot1 [\ldots]
\]

Furthermore, assume that we have some prior estimate that the center of the received valley should be $5$ bits from its actual position.

Then the received codeword would be as follows, where the underlined digit signifies our prior estimate for the center of the valley.

\[
w = [\ldots] 0000000000011111\underline{1}111111111111 [\ldots]
\]

Algorithm \ref{alg:align_valley} would start from this initial estimate and advance to the left, returning the correct center of the valley.

\[
w = [\ldots] 0000000000{\mathbf 0}111111111111111111 [\ldots]
\]

\begin{figure}
    \centering
    \includegraphics[width=0.9\columnwidth]{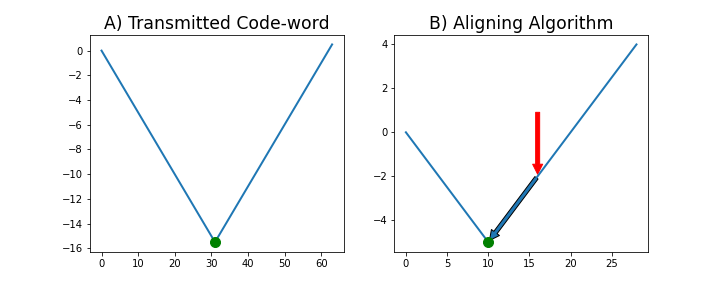}
    \caption{
    A visual representation of the algorithm for aligning valley centers on a specific example of a valley with parameters $(32, 32)$, being sent over a $\text{BDC}_p$ channel with parameter $p=\frac{1}{2}$.
    The plots go up by $\frac{1}{2}$ whenever the relevant string has a $1$ and down by $\frac{1}{2}$ whenever it has a $0$.
    {\textbf A)} The transmitted message, with the green dot representing the center of the valley.
    {\textbf B)} First, the channel deletes some of the bits resulting in a skewed valley. 
    We are given some prior estimate for the center of the valley (red arrow).
    The algorithm goes down the valley (blue arrow), until it terminates at the center (green dot).
    }
    \label{fig:align_alg}
\end{figure}

In Figure \ref{fig:align_alg} we show a geometric representation of this algorithm.

\section{Recursive Step}
\label{sec:recursion}
In this section we will define the recursive step in our construction, explain the rationale behind it and begin to prove its correctness. 
Intuitively, the main theorem we will prove here is that any error correcting code for the BDC with message length $k$ can be transformed into a code with message length $k^2$ and that the DFP, rate and complexity of the new code "scale well".
More formally we will show that:

\begin{theorem} [Recursive Step]
    \label{thm:rec_step}
    
    For some constants $c_1, c_2, c_3, k_0, \delta_0 > 0$ and for all $k > k_0$, $d > 0$, $2 ^ {k-1} - k > t > 0$, $\delta_0>\delta>0$, $1 > p > 0$ and any error correcting code $C$ for the $\text{BDC}_p$ channel with message length $k$, block length $n$ and DFP $\delta$, 
    
    there exists an error correcting code $C'$ for the $\text{BDC}_p$ with 
    message length $k^2$, 
    block length $n ^ \prime \leq(k+2t)(n+\frac{d}{1-p})$ and 
    DFP:
    \[
    \delta' \leq \Pr\left[\text{Bin}\left(\delta^{c_1}, k+2t\right) > t \right] + c_3 \left(k+2t\right) \exp{\left(-c_2\min{\{d,\frac{d^2}{k}\}}\right)}
    \]
    
    Furthermore, there exist an encoder and decoder for $C'$ with time complexity $\widetilde{O} \left(n ^ \prime\right)$ using up to $k+2t$ calls to the encoder and decoder of $C$.
\end{theorem}

We will use this construction in two scenarios: to improve the base of the recursion and for the steps of the recursion. 
In Table \ref{tab:recursive_thm}, we present the parameters of Theorem \ref{thm:rec_step} and the asymptotic values for both use cases.

\begin{table}[hbt!]
\centering
\begin{tabular}{c | c | c}

     \textbf{Parameter} & \textbf{Description} & \textbf{Recursion Base | Step} \\[1ex]
     \hline & \\
     $ k $ & Message Length & \\[1ex] \hline & \\
     $ n $ & Block Length & $\Theta\left(\frac{k}{1-p}\right)$ \\[1.5ex] \hline & \\
     $ d $ & Delimiter Length & $\Theta\left(k^{\frac{2}{3}}\right)$ \\[1.5ex] \hline & \\
     $ t $ & Reed Solomon Redundancy & $o\left(k\right)$ | $\Theta\left(k^ {\frac{2}{3}}\right)$ \\[1.5ex] \hline & \\
     $ \delta $ & Inner Code DFP & $o(1)$ | $\exp\left(-\frac{c_2}{2}k^{\frac{1}{6}}\right)$ \\[1.5ex] \hline & \\
      $c_1$ & Constant & $\frac{1}{34}$ \\[1.5ex] \hline & \\
      $c_2$ & Constant & $\frac{1}{256}$ \\[1.5ex] \hline & \\
      $c_2$ & Constant & $6$
      
\end{tabular}
\caption{\label{tab:recursive_thm} The parameters of Theorem \ref{thm:rec_step}, and their asymptotic values in the two use cases.}
\end{table}

The first set of values is used for improving of the base of the recursion.
In this case, our only bound on $\delta$ will be that it is an arbitrarily small constant, but its relation to $k$ will not be exactly known. 
Our goal in this step of the construction will be to reduce the error probability at the cost of an arbitrarily small but non-negligible cost to the rate of the code, and we will accomplish this by setting $t$  to be of the order of $\Theta\left(\delta^{\frac{c_1}{2}} k\right)$.

The second scenario is that of a step in our recursion. 
We will construct our recursion in such a manner that the DFP of the inner code will be bounded by $\delta < \exp{\left( -\frac{c_2}{2} k ^ {\frac{1}{6}} \right)}$. 
By setting $t = d = \Theta\left(k^{\frac{2}{3}}\right)$ we will be able ensure that on the one hand, the DFP of the final code will be $\delta' \ll \exp{\left(-\frac{c_2}{2} {\left(k ^ 2\right)} ^ {\frac{1}{6}}\right)}$ while on the other, the rate of the code will be reduced by only a factor of $1 - O\left(\frac{t+d}{k}\right) = 1 - O\left(k^{\frac{1}{3}}\right) = 1 - o(1)$.

The construction of the code $C': \{0, 1\} ^ {k^2} \rightarrow \{0, 1\} ^ {n ^ \prime}$ is as described in the introduction. 
First, the input string is split into $k$ parts of length $k$ each. 
These are considered as elements in an alphabet of size $2^k$ and a Reed Solomon encoding with parameters $\left[2^k, k+2t, k\right]$ is applied to them. 
Each of these is encoded using the encoder of $C$, a delimiter is appended to each codeword and the concatenation of all of these strings is outputted.

Similarly, the decoding algorithm works by locating the delimiters, separating them from the inner codewords and then decoding each inner codeword using the decoder of $C$. 
The decoded inner codewords are once again viewed as letters in an alphabet of size $2^k$ and the Reed Solomon decoding is applied.

We will define a delimiter with parameters $\alpha, \beta$ to be a valley of length $\beta$ surrounded by two valleys of length $\alpha$. 
In other words 

\[
\text{Delimiter}\left(\alpha, \beta\right) = \text{Valley}\left(\alpha\right) \text{Valley}\left(\beta\right) \text{Valley}\left(\alpha\right) = 0^\alpha 1^\alpha 0^\beta 1^\beta 0^\alpha 1^\alpha
\]

The exact values of $\alpha$ and $\beta$ will be discussed in Section \ref{sec:analysis}, but they will be of the order of $\alpha = \Theta\left(\frac{\log{\left(\frac{1}{\delta}\right)}}{1-p}\right)$ and $\beta = \frac{d}{2 (1 - p)} - 2\alpha = \Theta\left(\frac{k^{\frac{2}{3}}}{1-p}\right)$.

In order to complete the construction, we still need to provide methods of locating the delimiters in the received codeword and separating them from the inner codewords. These steps will be explained in the following subsections.

\subsection{Positioning Strings}
\label{subsec:positioning}
We will use the positioning strings to locate the delimiters one at a time.

Let us denote by $L_i$ the location of the center of the $i$th positioning string in the received codeword.
We can estimate the location of the center of the first positioning string as being around $\mathbb{E}\left[L_1\right] = (1-p) (n + 2\alpha + \beta)$ bits from the beginning of the received codeword. 
However, this is only a rough estimate and in reality $L_1 \sim  \text{Bin}\left(n + 2\alpha + \beta, 1-p\right)$.
We want to find the exact center.

This is where the valleys come into play.
With high probability, $\vert L_1 - \mathbb{E}\left[L_1\right] \vert$ will not be much larger than $\sigma_{L_1} = \sqrt{p (1 - p) (n + 2\alpha + \beta)}$ (where $\sigma_{L_1}$ is the standard deviation of $L_1$), and at least $\frac{1-p}{2} \beta$ of the bits on either side of the valley will survive the channel.
Therefore, so long as $\sqrt{p (1 - p) (n + 2\alpha + \beta)} \ll \frac{1-p}{2} \beta$, we can expect Algorithm \ref{alg:align_valley} to find the value of $L_1$ w.h.p.

Once we have found the $i$th delimiter, we go on to search for the $i+1$-th.
At each step we use the aligned center of the previous positioning string $L_i$ to obtain an estimate for $L_{i+1} \sim L_i + \text{Bin}\left(n + 4\alpha + 2\beta, 1-p\right)$, and use its valley to correct our estimate.
In Section \ref{sec:analysis} we will show that the probability that even a single positioning string will not be correctly found is at most $\left(k+2t\right) \exp{\left(-c_2\min{\{d,\frac{d^2}{k}\}}\right)}$

\subsection{Partitioning Strings}
\label{subsec:partitioning_string}

Once we have located the centers of all of the delimiters, we still need to separate them from the inner codewords.
This step is surprisingly difficult, because, unlike Con and Shpilka \cite{CS18} who designed their inner code to preserve a certain structure that would help them differentiate it from the delimiters, we reduce our overhead precisely by making no assumptions about the structure of the inner code.

For instance, if we were to use buffers of 0s as our delimiters and the inner codewords surrounding the delimiter happened to start / end with sequences of 0s, then we would have had a hard time telling which of the 0s belonged to the delimiter and which belonged to the inner codeword.
The same applies for any choice of the delimiters.

Our solution to this problem will be based on two ideas.
First, we will attempt to approximate the correct separation as accurately as possible.
However, this will not yield a perfect separation and we will need to mitigate the effects of this inaccuracy.

The first step will be accomplished by surrounding the positioning string with two valleys.
When decoding we can find the centers of these valleys by going from the positioning string until the ends of its valley and then proceeding to the bottom of the next/previous valleys.

Once we have found the center of a partitioning string, we can estimate the length of the faces of its valley.
Each of those will be i.i.d distributed according to $F \sim \text{Bin}\left(1-p, \alpha\right)$.
By guessing $f = (1-p) \alpha$ we will get a good approximation of $F$ (to within an error of $\Theta\left(\sqrt{p (1-p) \alpha}\right)$).

If we were to use this estimate as the separation between the delimiter and the inner codeword we would have a two sided error probability, either due to overshooting (attributing some bits of the inner codeword to the delimiter) or due to undershooting.

In the former case, this would cause us to delete some additional bits from the inner codeword.
Since the inner code is designed to deal with deletions, it stands to reason that it might also be able to decode received messages if they were also subject to a small amount of additional deletions.
This claim is not straightforward, since the inner code deals with random deletions and we will be subjecting it to a very structured set of deletions.
However, with a careful analysis we can bound the effect these deletions can have on the DFP of the inner code.

However, in the latter case we would end up erroneously inserting part of the delimiter to one end of the inner codeword.
Since the inner code is for a deletion channel, it might not be able to correctly decode the inner codeword, even after a small number of insertions.
Therefore, we have no way of bounding the effect this could have on its DFP.

That is why we want to reduce the probability of undershooting significantly more than the probability of overshooting.
To do this, instead of taking the estimate $f = \mathbb{E} F$, we will use the estimate $f = \mathbb{E} \left[F\right] + \eta \sigma_F$ for some parameter $\eta$ (where $\sigma_F$ is the standard deviation of $F$).

If $F$ was at most $\eta$ standard deviations from its expectancy, then we would have $f - F \in \left[0, 2\eta \sigma_F\right]$.
The lower bound means that no bits from the delimiter would ever trickle into the inner codeword, and the upper bound limits the number of bits from the inner codeword we will delete by accidentally attributing them to the delimiter.

In Section \ref{sec:analysis} we will bound both the probability that $f - F \notin \left[0, 2\eta \sigma_F\right]$ and the effect these deletions could have on the DFP of the inner code.

\section{The Recursive Construction}
\label{sec:base}

In the previous section we defined the manner in which we recursively concatenate our code to increase the message length at a negligible cost to the rate of the code and the complexity of its encoder and decoder.
In this section we will construct the base of this recursion and set the parameters for the recursive steps.

\subsection{The Base of the Recursion}
\label{subsec:base}
When choosing the base of the recursion we are in essence transforming a lower bound on the capacity of the channel to an actual error correcting code.
We will do this in a very inefficient manner, but as with Con and Shpilka's inner code, since this code will have message and block length of $O(1)$, this does not affect the asymptotic complexity of our construction.

In order to begin our recursion, we will need a base code with sufficiently low DFP and sufficiently large message length.
By definition, a lower bound on the capacity of the channel is a proof that there exists a family of codes for the channel with $r \geq \text{capacity} - \varepsilon$ for any $\varepsilon > 0$.
Therefore for any $k_0, \delta_0 > 0$ there are some codes in that family with message length $k > k_0$ and DFP $\delta < \delta_0$.
Let $\kappa$ be the smallest such block length. 
Since $\kappa$ is determined by $k_0$ and $\delta_0$, and since $k_0$ and $\delta_0$ are constant parameters of our construction, $\kappa = f\left(k_0, \delta_0\right) = O(1)$ must also be constant.

We will enumerate over values of $k > k_0$ and for each of them we will attempt to construct a base code.
This process will terminate when $k = \kappa$, so it will require only a finite number of iterations.

Since we are looking for an encoding map from some finite set of messages $\{0,1\}^k$ (where $k$ is the message length) to some finite set of codewords $\{0,1\}^n$ (where $n < \frac{9}{1-p} k = O(1)$ is the block length), and a decoding map from $\{0,1\}^{\leq n} \rightarrow \{0,1\}^k$, there are only finitely many pairs of this form.
By enumerating over all of these pairs and evaluating their DFP, we will be able to find a base code with message length $k$ if one exists. 

Since all of the steps in this process had a constant complexity, our construction of the inner code had a constant $O(1)$ complexity.
It should be noted that this algorithm is extremely inefficient and that we do not even know how to bound its complexity (except that it is $O(1)$).
We hope that future research will address this issue.

\subsection{Connecting the Recursive Steps}

All that remains now is to combine the recursive step shown in Section \ref{sec:recursion} with the base case constructed in the previous subsection.

Throughout most of the recursion we will use the recursive step defined in Theorem \ref{thm:rec_step} in the following setting:
\begin{equation}
    \begin{aligned}
    \delta_k &< \exp{\left( - \frac{c_2}{2} k ^ {\frac{1}{6}}\right)}\\
    d_k &= k^{\frac{2}{3}}\\
    t_k &= k^{\frac{2}{3}}
    \end{aligned}
    \label{eq:recursion_setting}
\end{equation}

Applying the recursion Theorem \ref{thm:rec_step}, for sufficiently large $k$, we have:

\begin{equation}
    \begin{aligned}
    \delta_{k^2} &\leq \Pr\left[\text{Bin}\left(\delta^{c_1}, k+2t\right) > t \right] + c_3 \left(k+2t\right) \exp{\left(-c_2\min{\{d,\frac{d^2}{k}\}}\right)}
    \end{aligned}
\end{equation}

We use Theorem \ref{thm:chernoff2} to bound the first term in the new decoding failure probability for sufficiently small $\delta$, by
\[
\Pr\left[\text{Bin}\left(\delta^{c_1}, k+2t\right) > t \right] \leq \exp\left( - \frac{c_1}{4} \log \left(\frac{1}{\delta}\right) t \right) \leq \frac{1}{2} \exp \left( - \frac{c_2}{2} t\right)
\]
and the second term for sufficiently large $k$ by
\[\left(k+2t\right) \exp{\left(-c_2\min{\{d,\frac{d^2}{k}\}}\right)} = \exp{\left( - c_2 k ^ {\frac{1}{3}} + O\left(\log{(k)}\right)\right)} \leq \frac{1}{2} \exp{\left( - \frac{c_2}{2} k ^ {\frac{1}{3}}\right)}
\].

Combining these inequality gives us a bound on the new DFP:

\begin{equation}
    \begin{aligned}
    \delta_{k^2} &\leq \exp{\left( - \frac{c_2}{2} k ^ {\frac{1}{3}}\right)}
    \end{aligned}
    \label{eq:recursive_DFP}
\end{equation}

The overhead of the code can also be easily bounded.
From the recursion theorem, we know that $n_{k^2} \leq \left(n_k + \frac{d_k}{1-p}\right) \left(k + 2 t_k\right)$.
Since $n_k \geq \frac{k}{1-p}$ (from the upper bounds on the capacity of these channels) and $d_k = t_k = k^{\frac{2}{3}}$, we have:

\begin{equation}
    \begin{aligned}
    n_{k^2} &\leq \left(1 + k^{-\frac{1}{3}}\right) ^ 2 \frac{k^2}{k} n_k \leq \left[\left(1 + k^{-\frac{1}{3}}\right) \cdot \left(1 + k^{-\frac{1}{6}}\right)\right] ^ 2  \frac{k^2}{\sqrt{k}} n_{\sqrt{k}} = \\
    & = \underbrace{\left[\left(1 + k^{-\frac{1}{3}}\right) \cdot \left(1 + k^{-\frac{1}{6}}\right) \cdot \ldots \cdot \left(1 + {\left(k_{\text{base}}\right)}^{-\frac{1}{3}} \right)\right] ^ 2}_{=: X} \frac{k^2}{k_{\text{base}}} n_{k_{\text{base}}} 
    \end{aligned}
    \label{eq:recursive_rate}
\end{equation}

Since $ n_{k_{\text{base}}} =  \frac{1}{r_{\text{base}}} k_{\text{base}}$ (where $r_{\text{base}}$ is the rate of the base code), it is easy to see that $r_{k^2} = \frac{k^2}{n_{k^2}} = \frac{1}{X} r_{\text{base}}$.
By bounding the value of $X$, we can bound the increase in the overhead of the code due to the recursion.

\begin{equation}
    \begin{aligned}
    X &= \left[\left(1 + k^{-\frac{1}{3}}\right) \cdot \left(1 + k^{-\frac{1}{6}}\right) \cdot \dots \cdot \left(1 + {\left(k_{\text{base}}\right)}^{-\frac{1}{3}} \right)\right] ^ 2 \\
    &\;\;\;\;\leq \exp{\left(2k^{-\frac{1}{3}} + 2k^{-\frac{1}{6}} + \dots + 2{\left(k_{\text{base}}\right)}^{-\frac{1}{3}}\right)} \leq \exp{\left(2\frac{{\left(k_{\text{base}}\right)}^{-\frac{1}{3}}}{1 - {\left(k_{\text{base}}\right)}^{-\frac{1}{3}}}\right)}
    \end{aligned}
    \label{eq:bounding_X}
\end{equation}

For a sufficiently large $k_{\text{base}}$, it is clear that this value can be set arbitrarily close to $1$, giving our code a nearly optimal rate.

However, this does not conclude our construction, since in each step of the recursion we assumed that $\delta_k \leq \exp{\left(-\frac{c_2}{2} k ^ {\frac{1}{3}}\right)}$, but our base construction only produced a code with an arbitrarily small $\delta_{\text{base}}$ and its relationship to $k_{\text{base}}$ is unknown.

\subsection{Completing the Construction}

In order to bridge this gap, we apply the recursive step to the base code one more time, with slightly different parameters.
We will denote by $k_0, \delta_0, d_0, t_0, n_0$ the parameters of the first application of the recursive step and by $k_{\text{base}} = {k_0}^2, n_{\text{base}}, \delta_{\text{base}}$ the parameters of the resulting code.

As before, we will set $d = k ^ {\frac{2}{3}}$, but unlike the previous setting, since $\delta_0$ is not necessarily as small as we would want it to be, we will need to set the value of $t$ to be somewhat larger.

In particular, we will set $t_0 = \ceil{{\delta_0} ^ {\frac{c_1}{2}} k_0}$.
Theorem \ref{thm:chernoff2} shows that, for sufficiently small $\delta_0$, the probability that a binomial variable with parameters $\left[\delta_0, k_0+2t_0\right]$ will be greater than $t_0$ is of the order of $\exp{\left(-\Theta(k_0)\right)}$.
Therefore, for sufficiently large $k_0$, the DFP after the first step of the recursion would be:

\begin{equation}
    \begin{aligned}
    \delta_{\text{base}} < \exp{\left(-\Theta(k_0)\right)} + \exp{\left( - c_2 {k_0} ^ {\frac{1}{3}} + o\left({k_0}^{\frac{1}{3}}\right)\right)} <  \exp{\left( - \frac{c_2}{2} {k_{\text{base}}} ^ {\frac{1}{6}}\right)}
    \end{aligned}
    \label{eq:base_DFP}
\end{equation}

Finally, we need to bound the rate of the entire code.
Using Equations \eqref{eq:recursive_rate} and \eqref{eq:bounding_X}, we are able to bound the rate:

\begin{equation}
    \begin{aligned}
    r_{k^2} &< X r_{\text{base}} \leq \exp{\left(2\frac{{\left(k_{\text{base}}\right)}^{-\frac{1}{3}}}{1 - {\left(k_{\text{base}}\right)}^{-\frac{1}{3}}}\right)} r_{\text{base}} \\
    &\;\;\;\;\leq\exp{\left(2\frac{{k_{0}}^{-\frac{1}{3}}}{1 - {k_{0}}^{-\frac{1}{3}}}\right)} \left(1 + {\delta_0} ^ {\frac{c_1}{2}} \right) r_{0}
    \end{aligned}
    \label{eq:final_rate}
\end{equation}

For sufficiently large $k_0$ and sufficiently small $\delta_0$, this can be arbitrarily close to $r_0$ which in turn can be arbitrarily close to the capacity of the channel, obtaining an arbitrarily close to optimal rate for our code.

\subsection{Decoding Complexity}

The recursive step promises at most $k+2t$ calls to the lower level of the construction and $\widetilde{O} \left(n ^ \prime\right)$ other operations.
Let $T_{\text{base}}$ be the encoding / decoding complexity of the base scenario, let $T_k$ be the total complexity of the operation for our code with message length $k$ and $I_k$ be the complexity due to operations which are not part of the lower levels of the recursion (i.e. finding the delimiters, separating them from the inner codewords and decoding the Reed Solomon encoding).

Similar to our bound on the rate of the code, we can bound the decoding / encoding complexities by:

\begin{equation}
    \begin{aligned}
    T_{k^2} &= (k+2t)T_k + I_{k^2} = I_{k^2} + \left(k+2k^{\frac{2}{3}}\right)T_k \\
    &= I_{k^2} + \left(k+2k^{\frac{2}{3}}\right)I_k + \left(k+2k^{\frac{2}{3}}\right)\left(\sqrt{k}+2k^{\frac{1}{3}}\right)T_{\sqrt{k}} \\
    & \leq I_{k^2} + \exp{\left(2\frac{{k_{0}}^{-\frac{1}{3}}}{1 - {k_{0}}^{-\frac{1}{3}}}\right)} \frac{k^2}{k} I_k + \left(k+2k^{\frac{2}{3}}\right)\left(\sqrt{k}+2k^{\frac{1}{3}}\right)T_{\sqrt{k}} \leq \dots\\
    & \dots\leq \exp{\left(2\frac{{k_{0}}^{-\frac{1}{3}}}{1 - {k_{0}}^{-\frac{1}{3}}}\right)} \left(\frac{k^2}{k} I_k + \frac{k^2}{\sqrt{k}} I_{\sqrt{k}} + \dots + \frac{k^2}{k_{\text{base}}} I_{k_\text{base}} + \frac{k^2}{k_{\text{base}}} T_{\text{base}}\right)
    \end{aligned}
    \label{eq:final_complexity}
\end{equation}

The recursive step can have at most a quasi-linear complexity on top of its calls to the inner code (see Theorem \ref{thm:rec_step}). 
Therefore, $I_k = \widetilde{O}\left(n\right)$.
Inserting this into Equation \eqref{eq:final_complexity}, we are can see that $T_{k^2} = \widetilde{O}\left(n^\prime\right)$.

\section{Adaptation to the Poisson Repeat Channel}
\label{sec:PRC}

In this section we will adapt the construction of our code for the BDC channel detailed in the last two sections, to the PRC channel.
Adapting the recursive step will be fairly straightforward.
However, adapting the base step will be a bit more tricky.

When working with the BDC, we used the fact that the received message could not be longer than the transmitted one.
This allowed us to build a decoding table that can return some value for any of the possible received messages.
However, the PRC could (with very low probability) expand a transmitted message to an arbitrarily long received message, and we will need to adapt our construction to address this issue.

\subsection{Adapting the Recursive Step}
\begin{theorem} [Recursive-Step for the PRC]
    \label{thm:rec_step_PRC}
    
    There exist some constants $c_1, c_2, k_0, \delta_0 > 0$, 
    such that for any $k > k_0$, $d > 0$, $2 ^ {k-1} - k > t > 0$, $\delta_0>\delta>0$, $\lambda > 0$ and any error correcting code $C$ for the $\text{PRC}_\lambda$ channel with block length $k$, message length $n$ and DFP $\delta$, 
    there exists an error correcting code $C'$ for the $\text{PRC}_\lambda$ with 
    block length $k^2$, 
    message length $n ^ \prime \leq(k+2t)(n+\frac{d}{\lambda})$ and 
    DFP $\delta' \leq \Pr\left[\text{Bin}\left(\delta^{c_1}, k+2t\right) > t \right] + \left(k+2t\right) \exp{\left(-c_2\min{\{d,\frac{d^2}{k}\}}\right)}$.
    
    Furthermore, there exist an encoder and decoder for $C'$ with time complexity $\widetilde{O} \left(n ^ \prime\right)$ using up to $k+2t$ calls to the encoder and decoder of $C$.
\end{theorem}

We will construct the recursive step almost exactly as in Section \ref{sec:recursion}, only changing $1-p$ to $\lambda$ in our conversion of lengths of bits over the channel.

\subsection{Adapting the Base of the Recursion}

In this section, we will adapt our construction of the base of our recursion from Section \ref{sec:base} to the PRC.
It is easy to see that for any $\lambda > 0$, there exists a family of codes for this channel with some non-negligible rate $\rho = \Theta(1)$ w.r.t the message length (for instance, by applying the jigsaw construction of \cite{MD06} with the Morse code distribution).
We will set $k_0, \delta_0$ to be the minimal value of the message length $k$ and DFP $\delta$ for which there exists such a base code that will suffice for our construction.

Unlike the previous construction, here we will only be able to approximate the DFP of our base code, so we will need to set two bounds.
Let $\kappa_1 \geq k_0$ be the minimal message length for which there is a code in the family of codes such that it has a DFP of at most $\delta_0$, and let $\kappa_2 \geq k_0$ be the minimal message length for which there is a code in the family of codes such that it has a DFP of at most $\frac{\delta_0}{2}$ .
Similar to the construction in Section \ref{subsec:base}, we do not have an explicit bound on $\kappa_1, \kappa_2$, but we know that they are bounded by some $O(1)$ constant.

As in Section \ref{subsec:base}, we will enumerate over values of $k \geq k_0$, but this time we only promise that our enumeration will end somewhere between $\kappa_1$ and $\kappa_2$.
For each such $k$, we enumerate over all $n\leq \rho k$, encoders $C:\left\{0,1\right\}^k \rightarrow \left\{0,1\right\}^n$ and decoders $D:\left\{0,1\right\}^m \rightarrow \left\{0,1\right\}^k$, where $m$ is the smallest integer for which:
\[
\Pr\left[\text{Poisson}(\lambda n) \geq m\right] \leq \frac{\delta}{2}
\]
For each of these encoder-decoder pairs, we enumerate over all messages in $\left\{0,1\right\}^k$ and encode them using the encoder.
For each codeword, we enumerate over all possible results of applying the channel to the codeword that have output length at most $m$.

We sum the probabilities of the eventualities where this process would result in a decoding failure (cases where the channel outputted more than $m$ bits are counted as failures), and take the message with the highest DFP.
This gives us an approximation of 
\[
\text{DFP}_{\text{actual}} \leq \text{DFP}_{\text{estimate}} \leq \text{DFP}_{\text{actual}} + \frac{\delta}{2}
\]

If the estimated DFP is at most $\delta$, then we output the pair of encoder-decoder tables.
It is easy to see that if the actual DFP is at most $\frac{\delta}{2}$ then the estimate DFP is at most $\delta$ and we will output it.
Therefore our process either terminates before $\kappa_2$ or at $\kappa_2$ and must have a constant complexity.

\section{Analysis of Decoding Failures}
\label{sec:analysis}

In order to complete our construction, we still need to show that the recursive construction of the code for the $\text{BDC}_p$ channel in Section \ref{sec:recursion} and its adaptation to the $\text{PRC}_\lambda$ in Section \ref{sec:PRC}, do indeed have a low DFP.
There are three main steps in the decoding process, and we will need to bound the failure probability of each of them.

The first step is locating all of the positioning strings.
When this is complete, we will separate the inner codewords from the delimiters and decode each one using the inner code.
Finally, we use the Reed-Solomon decoding to return the original message.

\subsection{Outer-Code Failure}

\subsubsection{For the Binary Deletion Channel}

We begin with analysing the probability that we will fail to find at least one of the delimiters for the $\text{BDC}_p$ channel.

Let $F_i$ denote the event that we failed to align the $i$th positioning string.
Since we locate the positioning strings serially, we will begin by bounding the probability that we failed to locate the $i$th positioning string, given that we did align the previous $i-1$ delimiters successfully.
In other words, we want to bound $\Pr{\left[F_i \mid \overline{F_1}, \dots, \overline{F_{i-1}}\right]}$.

The algorithm we use to find the center of the $i$th positioning string in the received message $L_i$ (as discussed in Section \ref{subsec:positioning}), is as follows:

First, we use the location of the previous positioning string $L_{i-1}$ to obtain an estimate for $L_i$, using the fact that $L_i - L_{i-1} \sim \text{Bin}\left(1-p, n + 4\alpha + 2\beta\right)$.
So we set our initial estimate $G_i = L_{i-1} + (1-p) \left(n + 4\alpha + 2\beta\right)$.

In order for our algorithm to succeed, this estimate must be within the corresponding valley in the received codeword.
The length of either face of the received valley is distributed according to $S\sim \text{Bin}\left(1-p, \beta\right)$, so according to Theorem \ref{thm:chernoff}, both faces are of length at least $\frac{(1-p)\beta}{2}$ w.p. $\geq 1 - 4\exp{\left(- \frac{1}{16} d\right)}$.

The probability that our estimate was off by more than $\frac{(1-p)\beta}{2}$ can be bounded using the Chernoff bound:

\begin{equation}
    \begin{aligned}
    \Pr{\left[ \vert G_i -  L_{i-1} - (1-p) \left(n + 4\alpha + 2\beta\right)\vert \geq \frac{(1-p)\beta}{2}\right]} &\leq 2\exp{\left( - \frac{1}{4} \left( \frac{d}{2(k+d)} \right) ^ 2 \right)} \\
    &\leq 2\exp{\left( - \frac{1}{256} \frac{d ^ 2}{k} \right)}
    \end{aligned}
\end{equation}

Using the union bound we get:

\begin{equation}
    \begin{aligned}
    \Pr{\left[F_i \mid  \overline{F_1}, \dots, \overline{F_{i-1}} \right]} &\leq  4\exp{\left(- \frac{1}{32} d \right)} + 2\exp{\left( - \frac{1}{256} \frac{d ^ 2}{k} \right)} \leq \\
    &\leq 6\exp{\left( - \frac{1}{256} \min\left\{\frac{d ^ 2}{k}, d\right\} \right)}
    \end{aligned}
    \label{eq:BDC_F_i}
\end{equation}

\subsubsection{For the Poisson Repeat Channel}
Similarly, for the $\text{PRC}_\lambda$ , we wish to bound $\Pr{\left[F_i \mid \overline{F_1}, \dots, \overline{F_{i-1}}\right]}$.

This time $L_i - L_{i-1} \sim \text{Poisson}\left(\lambda\left(n + 4\alpha + 2\beta\right)\right)$ and our estimate will be $G_i = L_{i-1} + \lambda \left(n + 4\alpha + 2\beta\right)$.
The length of the side of each received valley is now distributed according to $S\sim \text{Bin}\left(\lambda \beta\right)$ and the probability that it would be less than $\frac{\lambda}{2}\beta$ is bounded by $4\exp{\left(- \frac{1}{16} d\right)}$ (see Theorem \ref{thm:chernoff}).

We will use the Chernoff bound (Theorem \ref{thm:chernoff}) to show that:
\begin{equation}
    \begin{aligned}
    \Pr{\left[ \left\vert G_i -  L_{i-1} - \lambda \left(n + 4\alpha + 2\beta\right) \right\vert \geq \frac{\lambda\beta}{2}\right]} &\leq 2\exp{\left( - \frac{1}{4} \left( \frac{d}{2(k+d)} \right) ^ 2 \right)} \\
    &\leq 6\exp{\left( - \frac{1}{256} \min\left\{\frac{d ^ 2}{k}, d\right\} \right)}
    \end{aligned}
\end{equation}

Applying the union bound once more, we see that Equation \eqref{eq:BDC_F_i} holds for our $\text{PRC}_\lambda$ code as well.

\subsubsection{Combining the Terms}

So far, we have bounded the terms $\Pr{\left[F_i \mid  \overline{F_1}, \dots, \overline{F_{i-1}} \right]}$, but what we really want is to show that $\Pr{\left[F_1 \vee \dots \vee F_{k+2t} \right]}$ is negligible.
Indeed:

\begin{equation}
    \begin{aligned}
    \Pr{\left[F_1 \vee \dots \vee F_{k+2t} \right]} &= 1 - \Pr{\left[\overline{F_1} \bigwedge \dots \bigwedge \overline{F_{k+2t}} \right]} =\\
    &=1 - \prod_i \left(1 - \Pr{\left[F_i \mid  \overline{F_1}, \dots, \overline{F_{i-1}} \right]}\right) \\
    &\leq \sum_i \Pr{\left[F_i \mid  \overline{F_1}, \dots, \overline{F_{i-1}} \right]}
    \end{aligned}
\end{equation}

Note that this inequality is not the union bound, since the right-hand-side is of the form $ \sum_i \Pr{\left[F_i \mid  \overline{F_1}, \dots, \overline{F_{i-1}} \right]} $ (and not $ \sum_i \Pr{\left[F_i\right]} $). This proves that the probability of missing even a single delimiter is bounded by the second term of the new DFP in Theorem \ref{thm:rec_step}.

\subsection{Inner-Code Failure}

Suppose we have succeeded in finding the centers of all of the positioning strings.
The remaining steps in the decoding algorithm are:

\begin{itemize}
    \item Use the partitioning strings to separate the inner codewords from the delimiters.
    \item Decode each of the inner codewords using the recursive decoder.
    \item Use the redundancy of the Reed-Solomon to correct up to $t$ failures of the inner code.
\end{itemize}

Note that the first two types of errors are local in the sense that any delimiter that mixes in with an inner codeword and any inner codeword that is incorrectly decoded does this independently from the rest of the codeword.
This is in contrast to the third step, which is global in the sense that for it to succeed we need to make sure that enough of the local steps are successful.
We will begin by bounding the probability of any single local step failing, and then we will use their independence to bound the probability that many of them will fail.

\subsubsection{Delimiter Estimation}

There are many types of errors that can occur when trying to decode the inner codeword between two positioning strings.
We divide them into those we attribute to the delimiter estimation and those we attribute to the inner code decoding.

We will say that the delimiter estimation failed if we either missed the delimiter completely or underestimated the length of the delimiter (effectively inserting new bits into the received inner codeword).
We will say that the inner decoding failed if the inner message was incorrectly decoded, despite a successful delimiter estimation.

\paragraph{The Probability of Missing a Delimiter}
A necessary condition for us to completely miss a delimiter is that at least one side of its valley was completely deleted by the channel.
We set the size of the positioning strings to be $\alpha = -C \frac{\log\left(\delta\right)}{1-p}$ (or $\alpha = -C \frac{\log\left(\delta\right)}{\lambda}$) for some positive constant $C$.
Since either side of the valley is of length $\alpha=\Theta(-\log(\delta))$ before the channel, the probability that it will be completely deleted is $p^\alpha\leq\delta^{C}$ for the BDC and $\exp\left(-\lambda\alpha\right) \leq \delta^{C}$ for the PRC.

\paragraph{The Probability of Under-Estimating a Delimiter}

Let $F = \sum_{1 \leq i \leq \alpha} x_i$ be the length of the face of the valley separating between the center of the partitioning string and the inner codeword (after the noise of the channel).
In the case of the BDC each of the $x_i$ is i.i.d $\text{Bernoulli}(1-p)$ and $\mathbb{E} F = (1-p) \alpha $ and in the case of the PRC each of the $x_i$ are i.i.d $\text{Poisson}(\lambda)$ and $\mathbb{E} F = \lambda \alpha $.

Our strategy for this part of the algorithm is to try to obtain the smallest overestimation of the value of $F$, while minimizing the probability of an underestimation.
Setting our estimate for $F$ to be $f_{\text{estimate}} = 2\mathbb{E} F$ works well for our purposes.

In order to quantify this statement, we need to show that the probability of underestimating is small and bound the effect such an overestimation could have on the DFP of the inner code.
For the former, a simple application of the Chernoff bound (Theorem \ref{thm:chernoff}) shows that for either model, we have:

\begin{itemize}
    \item In the BDC model $\Pr\left[ F > f_{\text{estimate}} \right] \leq 2 \exp\left( -\frac{1}{16} \alpha (1-p) \right) \leq 2 \delta^{\frac{C}{16}}$
    \item In the PRC model $\Pr\left[ F > f_{\text{estimate}} \right] \leq 2 \exp\left( -\frac{1}{16} \alpha \lambda \right) \leq 2 \delta^{\frac{C}{16}}$
\end{itemize}

In either case, $\Pr\left[F > f_{\text{estimate}} \right] \leq 2\delta ^ {\frac{C}{16}}$.

\subsubsection{The DFP with Additional Deletions}

Finally, we need to bound the probability that we will fail to decode the inner codeword -- possibly due to an over-estimation of $F$.
We will do this with a method similar to Bayesian analysis.

Let $\Omega = \mathbb{N} ^ n$ be the sample space whose members represent possible values for the number of repetitions of each bit in the inner codeword.
For instance, in the $\text{BDC}_p$ model, only samples of the form $\vec{e} \in \left\{0, 1\right\} ^ n \subseteq \Omega$ can occur, and they have a probability of

\[
\Pr\left[ \vec{e} \mid \text{BDC}_p \right] = p^{n - \left\vert e \right\vert} \left(1 - p\right)^{n - \left\vert e \right\vert}
\]

Let $C$ be a channel ($C \in \left\{ \text{BDC}_p, \text{PRC}_\lambda \right\}$), and let $E$ and $D$ be the internal encoding and decoding algorithms.
We will denote by $C + \text{Overestimation}_{\ell_1,\ell_2}$ the noise model obtained by applying the channel $C$, followed by a deletion of the first and last $\ell_1, \ell_2$ bits of the received message (due to an overestimation of $F$).

Furthermore, let $m\in \left\{0, 1\right\} ^ k$ be some inner message, and let $B_m \subseteq\Omega$ be the set of channels noises for which the decoding of $m$ will fail.
Our recursive step requires that:

\[
\max_m \left\{\Pr\left[ B \mid C \right]\right\} = \text{DFP}(E,D,C) = \delta
\]

If we were able to show a bound of the form 
\[
\forall e \in \Omega \;\; \Pr\left[ \vec{e} \mid C + \text{Overestimation}_{\ell_1,\ell_2} \right] \leq \zeta \Pr\left[ \vec{e} \mid C \right]
\]
then it would allow us to prove a bound of the form:

\begin{equation}
    \begin{aligned}
    \text{DFP}\left(E,D,C + \text{Overestimation}_{\ell_1,\ell_2}\right) &= \max_m \left\{\Pr\left[ B \mid C + \text{Overestimation}_{\ell_1,\ell_2} \right]\right\} \\
    &\leq \zeta \max_m \left\{\Pr\left[ B \mid C\right]\right\} = \zeta \delta
    \end{aligned}
\end{equation}

However, we cannot prove this bound for a sufficiently small $\zeta$.
Consider the simple case of a $C = \text{BDC}_p$, $\ell_1 = 1$, $\ell_2 = 0$ and $p=\frac{1}{2}$.
We would hope to have a bound for this case with $\zeta = \Theta(1)$.
For most common cases of $e$, such a bound would hold. However, when $\vec{e} = \vec{0}$, then:

\[
\Pr\left[ \vec{e} \mid C \right] = 2 ^ {-n}
\]

\[
\Pr\left[ \vec{e} \mid C + \text{Overestimation}_{\ell_1,\ell_2} \right] = \left( n + 1\right) 2^{-n}
\]

This combination implies that $\zeta > n + 1 = \omega(1)$, and this will not be sufficiently small for some of the cases of our recursive step.
However, this example is a fairly extreme scenario.
This is not a coincidence and we will prove that aside from a few very rare cases, we can indeed show a stronger bound.

Specifically, we will split the effects of the channel into two categories.
We will say that a set of deletions of the channel is a member of the {\em common case}, if of the first and last $4\alpha$ bits of the inner codeword, at least $2(1-p)\alpha$ (or $2 \lambda \alpha$ in the PRC) remained after the deletions of the channel.
Otherwise, we will say that this set of deletions is a member of the {\em extreme case}.

We will then prove bounds of the form:

\[
\forall \vec{e} \in \text{Common Case}\;\;\Pr\left[ \vec{e} \mid C + \text{Overestimation}_{\ell_1,\ell_2} \right] \leq \zeta_{\text{common}} \Pr\left[ \vec{e} \mid C \right]
\]

\[
\Pr\left[ \overline{\text{Common Case}} \mid C + \text{Overestimation}_{\ell_1,\ell_2} \right] \leq \delta_{\text{extreme}}
\]

Combining these inequalities we will get the following inequality:

\begin{equation}
    \begin{aligned}
    &\text{DFP}\left(E,D,C + \text{Overestimation}_{\ell_1,\ell_2}\right) = \max_m \left\{\Pr\left[ B \mid C + \text{Overestimation}_{\ell_1,\ell_2} \right]\right\} \\
    &\;\;\;\;\leq \delta_{\text{extreme}} + \zeta_{\text{common}} \max_m \left\{\Pr\left[ B \mid C\right]\right\} = \delta_{\text{extreme}} + \zeta_{\text{common}} \delta
    \label{eq:combined_DFP}
    \end{aligned}
\end{equation}

\paragraph{Bounding $\zeta_{\text{common}}$}

In the common case, the first and last $2 (1-p) \alpha$ (or $2 \lambda \alpha$) bits of the received inner message correspond to at most $4\alpha$ of the first and last bits of the encoded inner message.

Since $F \geq 0$, our overestimation could have deleted at most $\ell_i = f_{\text{estimate}} - F \leq f_{\text{estimate}} = 2 (1-p) \alpha$ (or $2 \lambda \alpha$) bits from the start / end of the received message.
Denote by $t_1 / t_2$ the number of bits from the start/end of the transmitted inner codeword that correspond to the first/last $\ell_1 / \ell_2$ bits of the received inner codeword.
Since we are in the common case, these are bounded by $t_i \leq 4 \alpha$.

Since the first/last $t_1 / t_2$ bits of the transmitted message could have been deleted by the channel w.p. $p^{t_i}$ (or $\exp\left(-t_i \lambda\right)$) anyways, and since this is independent of the rest of the deletions or repetitions of the channel, we can obtain the following bounds on $\zeta_{\text{common}}$ for $\text{BDC}_p$ and for the $\text{PRC}_\lambda$:

\begin{equation}
    \begin{aligned}
        &\zeta_{\text{common}}\left(\text{BDC}_p\right) \leq  \frac{1}{\Pr\left[\text{first and last bits } t_1,\;t_2 \text{ were deleted} \mid \text{BDC}_p\right]} \\ 
        &\;\;\;\;\;\;\;\;\;\;\;\;\;\;\;\;\;\;\;\;\;\;\;\;\;\;\;\;\;\;\;\;\leq p^{-t_1} p^{-t_2} \leq p^{-8\alpha} \leq \delta^{-8C}\\
        &\zeta_{\text{common}}\left(\text{PRC}_\lambda\right) \leq  \frac{1}{\Pr\left[\text{first and last bits } 4\alpha \text{ were deleted} \mid \text{PRC}_\lambda\right]} \\ 
        &\;\;\;\;\;\;\;\;\;\;\;\;\;\;\;\;\;\;\;\;\;\;\;\;\;\;\;\;\;\;\;\;\leq \exp\left( \left(t_1+t_2\right)\lambda \right)\leq \exp\left( 8\alpha\lambda \right) \leq \delta^{-8C}
        \label{eq:zeta_common}
    \end{aligned}
\end{equation}

\paragraph{Bounding $\delta_{\text{extreme}}$}

The extreme case is defined as the scenario where only $\frac{1-p}{2}$ (or $\frac{\lambda}{2}$) of the first or last $4\alpha$ bits of the transmitted inner message survived the channel.

Using the Chernoff bound (Theorem \ref{thm:chernoff}), we can easily see that this probability is bounded by:

\begin{equation}
    \begin{aligned}
        &\delta_{\text{extreme}}\left(\text{BDC}_p\right) \leq  \Pr \left[ \text{Bin}(1-p, 4\alpha) < 2(1-p) \alpha\right] \leq 2\exp \left(- \frac{1}{4} \alpha (1-p) \right) \leq 2\delta^{\frac{C}{4}}\\
        &\delta_{\text{extreme}}\left(\text{PRC}_\lambda\right) \leq  \Pr \left[ \text{Poisson}(\lambda, 4\alpha) < 2\lambda \alpha\right] \leq 2\exp \left(- \frac{1}{4} \alpha \lambda \right) \leq 2\delta^{\frac{C}{4}}
        \label{eq:delta_extreme}
    \end{aligned}
\end{equation}

\paragraph{Combining these Bounds}

Combining Equations \eqref{eq:zeta_common} and \eqref{eq:delta_extreme} with Equation \eqref{eq:combined_DFP} we can bound the probability that any single inner codeword will be incorrectly decoded by:

\begin{equation*}
    \begin{aligned}
    &\Pr\left[ \text{Decoding Failure} \mid C + \text{Delimiter Decoding} \right] \\
    &\;\;\;\;\leq \zeta_{\text{common}} \delta + \delta_{\text{extreme}} + \Pr\left[ \text{Delimiter Estimation Failure}\right] \\
    &\;\;\;\;\leq\delta ^ {1 - 8C} + 2 \delta^{\frac{C}{4}} + \delta ^ C
    \end{aligned}
\end{equation*}

Setting $C = \frac{4}{33}$ we can bound the final DFP by:

\begin{equation}
    \begin{aligned}
        \Pr\left[ \text{Decoding Failure} \mid C + \text{Delimiter Decoding} \right] \leq 4 \delta ^ {\frac{1}{33}}
        \label{eq:final_inner_DFP}
    \end{aligned}
\end{equation}

For sufficiently small $\delta$, this is bounded by $\delta ^ {\frac{1}{34}}$.

\subsubsection{Reed Solomon Decoding}

In the previous subsections we showed that the probability that we will fail to decode any single inner codeword is $\varepsilon \leq \delta ^ {\frac{1}{34}}$.
Since each of the decoding process of each inner codeword is independent of the other inner codewords, the number of inner decoding processes that fail is distributed according to $\text{Bin}\left(\varepsilon, k+2t\right)$.

Since $f(\varepsilon) = \Pr\left[\text{Bin}\left(\varepsilon, k+2t\right) > t\right]$ is a monotone function, it is clear that the probability of a decoding failure due to inner code failures is bounded by the first term in the DFP of Theorem \ref{thm:rec_step}:

\[
\Pr\left[ \text{Reed Solomon Error} \right] = \Pr\left[\text{Bin}\left(\varepsilon, k+2t\right) > t\right] \leq \Pr\left[\text{Bin}\left(\delta ^ {\frac{1}{34}}, k+2t\right) > t\right]
\]

This concludes the proof of Theorems \ref{thm:rec_step} and \ref{thm:rec_step_PRC} for $c_1 = \frac{1}{34}$, $c_2 = \frac{1}{256}$ and $c_3 = 6$.

\section{Discussion}
\label{sec:discussion}

In this paper we presented several new techniques for constructing error-correcting--codes for asynchronous channels, and used them to prove a method of transforming lower bounds on the capacities of the BDC and PRC channels to efficient encoding and decoding algorithms.
This answers the main question considered \cite{CS18} and \cite{guruswami2018polynomial}.

However, this construction is far from practical.
For instance, the inner code at the basis of our recursion is constructed in a doubly exponential time in the size of the inner codes message length.
Even our bound on the decoding failure of the inner code is  $\delta ^ {\frac{1}{34}} = \exp\left( \frac{k^{\frac{1}{6}}}{34} \right)$ which is technically $o(1)$ but converges extremely slowly.

Furthermore, this work deals only with BDC and PRC channels.
While these channels offer us a chance to model the behaviour of asynchronous channels, they do not represent the more realistic channels which contain both synchronisation errors and bit-flipping errors.
To this end, the binary InsDel channel offers a more comprehensive model and it remains an open question whether the methods described here can be used to construct efficient codes for it as well.

Finally, now that we have a framework for efficient encoding and decoding for the BDC and PRC channels, we can return to the question of the capacity of these channels.
We know that when $p\rightarrow 1$ this capacity scales proportionally to $1-p$, but the factor of this conversion is still unknown.
Mitzenmacher and Drinea \cite{MD06} showed that these capacities are at least $\frac{1-p}{9}$, and \cite{dalai2011new} gave an upper bound of $0.4143 (1 - p) \pm o_n (1)$, but this gap is far from closed.

\section*{Acknowledgements}

We thank Roni Con, Aviad Rubinstein and Muli Safra for their helpful comments on previous drafts. This works was partially funded by the Deutsch institute fund.

\bibliographystyle{alpha}
\bibliography{main}

\section{Appendix}
\subsection{Proof of Theorem \ref{thm:chernoff}}
\label{app:chernoff}
The Chernoff bound for binomial distributions shown in \cite{boucheron2013concentration} is:

\begin{equation}
    \begin{aligned}
    \Pr \left[ \left\vert\text{Bin}\left(p, n\right) - p n \right\vert > \varepsilon p n \right] &\leq
    \Pr \left[ \text{Bin}\left(p, n\right) - p n > \varepsilon p n \right] +
    \Pr \left[ \text{Bin}\left(p, n\right) - p n  < - \varepsilon p n \right] \\
    &\leq \exp\left(-h_p\left((1+\varepsilon) p \right)\right) + \exp\left(-h_{1-p}\left(1 - (1-\varepsilon) p \right)\right)
    \label{eq:clean_chernoff}
    \end{aligned}
\end{equation}

Where $ h_p (x) = (1-x) \log \left(\frac{1-x}{1-p}\right) + x \log \left(\frac{x}{p}\right)$.
It is easy to see that:

\begin{equation*}
    \begin{aligned}
    & \frac{d}{dx} h_p (x) = \log \left(\frac{1-x}{1-p}\right) + \log \left(\frac{x}{p}\right)\\
    & \frac{d^2}{dx^2} h_p (x) = \frac{1}{x-x^2}
    \end{aligned}
\end{equation*}

When setting $x = (1 \pm \varepsilon) p$, we get the following inequality:
\begin{equation*}
    \begin{aligned}
    & h_p \left((1 + \varepsilon) p\right) = h_p (p) + h^{\prime}_p (p) \varepsilon p + \int_{p < x < (1+\varepsilon) p} \int_{p < \xi < x} h^{\prime \prime}_p (\xi) \geq \frac{1}{4} \varepsilon ^ 2 p \\
    & h_p \left((1 - \varepsilon) p\right) = h_p (p) - h^{\prime}_p (p) \varepsilon p + \int_{(1+\varepsilon) p < x <  p} \int_{p < \xi < x} h^{\prime \prime}_p (\xi) \geq \frac{1}{4} \varepsilon ^ 2 p
    \end{aligned}
\end{equation*}

Plugging this into Equation \eqref{eq:clean_chernoff}, we prove the first case of Theorem \ref{thm:chernoff}:

\begin{equation}
    \begin{aligned}
    \Pr \left[ \left\vert\text{Bin}\left(p, n\right) - p n \right\vert > \varepsilon p n \right] \leq 2 \exp \left( - \frac{1}{4} \varepsilon ^ 2 p n \right)
    \end{aligned}
\end{equation}

The Chernoff bound for Poisson distributions shown in \cite{boucheron2013concentration} is:

\begin{equation}
    \begin{aligned}
    \Pr \left[ \left\vert\text{Poisson}\left(\lambda\right) - \lambda \right\vert > \varepsilon \lambda \right] &\leq
    \Pr \left[ \text{Poisson}\left(\lambda\right) - \lambda > \varepsilon \lambda \right] +
    \Pr \left[ \text{Poisson}\left(\lambda\right) - \lambda  < - \varepsilon \lambda \right] \\
    &\leq \exp\left( -\lambda h(\varepsilon) \right) + \exp\left(-\lambda h(-\varepsilon)\right)
    \label{eq:poisson_chernoff}
    \end{aligned}
\end{equation}

Where $h(x) = (1+x) \log \left(1+x\right) - x$.
As with the binomial distribution, we will consider the derivatives of $h$:

\begin{equation*}
    \begin{aligned}
    & \frac{d}{dx} h (x) = \log \left(1-x\right)\\
    & \frac{d^2}{dx^2} h (x) = \frac{1}{1+x}
    \end{aligned}
\end{equation*}

Therefore:
\begin{equation*}
    \begin{aligned}
    & h \left(\varepsilon\right) = h (0) + h^{\prime} (0) \varepsilon p + \int_{0 < x < \varepsilon} \int_{0 < \xi < x} h^{\prime \prime} (\xi) \geq \frac{1}{4} \varepsilon ^ 2 \lambda \\
    &  h \left(-\varepsilon\right) = h (0) - h^{\prime} (0) \varepsilon p + \int_{0 < x < \varepsilon} \int_{0 < \xi < x} h^{\prime \prime} (\xi) \geq \frac{1}{4} \varepsilon ^ 2 \lambda
    \end{aligned}
\end{equation*}

Plugging this into Equation \eqref{eq:poisson_chernoff}, we prove the second case of Theorem \ref{thm:chernoff}:

\begin{equation}
    \begin{aligned}
    \Pr \left[ \left\vert\text{Poisson}\left(\lambda\right) - \lambda \right\vert > \varepsilon \lambda \right] \leq 2 \exp \left( - \frac{1}{4} \varepsilon ^ 2 \lambda\right)
    \end{aligned}
\end{equation}

\subsection{Proof of Theorem \ref{thm:chernoff2}}
\label{app:chernoff2}

We recall once more the Chernoff bound from \cite{boucheron2013concentration}

\[
    \Pr \left[ \text{Bin}\left(p, n\right) > p n + a n\right] \leq \exp \left( - h_p (a) n \right)
\]

We can bound this value through a simple analysis of $h_p (a)$ when $a = \alpha p$ for some $\alpha > e^2$, concluding the proof.

\begin{equation*}
    \begin{aligned}
    h_p (a) = (1-a) \log \left(\frac{1-a}{1-p}\right) + a \log \left(\frac{a}{p}\right) \geq -a + a \log (\alpha) \geq \frac{1}{2} a \log (\alpha)
    \end{aligned}
\end{equation*}

\end{document}

%% file: preamble.tex
\usepackage[utf8]{inputenc}
\usepackage{hyperref}
\usepackage{color}
\usepackage{graphicx}
\graphicspath{ {./Images/} }
\usepackage{wrapfig}
\usepackage{amsmath, amsthm, amssymb}
\usepackage{theoremref}

\newtheorem{theorem}{Theorem}[section]

   \newtheorem{definition}{Definition}

\newcommand\ceil[1]{\lceil{#1}\rceil}

\usepackage{soul}
\usepackage{xcolor}

\usepackage{amsmath}
\makeatletter
\def\resetMathstrut@{%
  \setbox\z@\hbox{%
    \mathchardef\@tempa\mathcode`\[\relax
    \def\@tempb##1"##2##3{\the\textfont"##3\char"}%
    \expandafter\@tempb\meaning\@tempa \relax
  }%
  \ht\Mathstrutbox@\ht\z@ \dp\Mathstrutbox@\dp\z@}
\makeatother